\documentclass[aps,preprint,superscriptaddress]{revtex4}
\usepackage{amssymb}
\usepackage{amsmath}
\usepackage{graphicx}
\usepackage{fancyhdr}
\newcommand{\bea}{\begin{eqnarray}}
\newcommand{\eea}{\end{eqnarray}}

\begin{document}

\title{KeV right-handed neutrinos from type II seesaw mechanism in a 3-3-1 model}

\author{D. Cogollo}
\email{diegocogollo@fisica.ufpb.br}
\affiliation{{ Departamento de
F\'{\i}sica, Universidade Federal da Para\'\i ba, Caixa Postal 5008, 58051-970,
Jo\~ao Pessoa, PB, Brasil}}

\author{H. Diniz}
\email{hermes@fisica.ufpb.br}
\affiliation{{ Departamento de
F\'{\i}sica, Universidade Federal da Para\'\i ba, Caixa Postal 5008, 58051-970,
Jo\~ao Pessoa, PB, Brasil}}

\author{C.A. de S. Pires}
\email{cpires@fisica.ufpb.br}
\affiliation{{ Departamento de
F\'{\i}sica, Universidade Federal da Para\'\i ba, Caixa Postal 5008, 58051-970,
Jo\~ao Pessoa, PB, Brasil}}

\date{\today}

\begin{abstract}

We adapt the type II seesaw mechanism to the framework of the  3-3-1 model with right-handed neutrinos. We emphasize that the mechanism is capable of generating small masses for the  left-handed and right-handed neutrinos and  the structure of the model allows that both masses arise from the same  Yukawa coupling. For typical values of the free parameters of the model we may obtain at least one right-handed neutrino with mass in the KeV range.  Right-handed neutrino with mass in this range is a viable candidate for the warm component of the dark matter existent in the universe.

\end{abstract}

\pacs{}
\maketitle

\section{Introduction}
\label{sec1}
One of the main puzzles in particle physics concerns the smallness of the neutrino masses. It is considered that the most elegant way of generating  small Majorana neutrino masses is through seesaw mechanisms. There are three distinct ways of accomplishing the seesaw mechanism into the standard model. In the so-called type I seesaw mechanism\cite{typeI}, small neutrino masses are induced by three heavy right-handed neutrinos, while in the type II\cite{typeII} and type III\cite{typeIII} seesaw mechanisms, small neutrino masses are induced by a heavy triplet of scalars and a heavy triplet of leptons, respectively. All these mechanisms were originally developed to induce small masses to the left-handed neutrinos.

Right-handed neutrinos were not detected yet in nature. Nobody knows if they are light or heavy particles. Light right-handed neutrinos are phenomenologicaly interesting  because of their intricated implications in particle physics\cite{Part-phys}, astrophysics and cosmology\cite{astro-cosmo}. For example, warm dark matter in the form of sterile neutrinos with mass in the KeV range has been advocated as a solution to  the conflict among cold dark matter and observations of clustering on subgalactic scales\cite{bode}. There are many papers devoted to the study of such implications\cite{dodelson,implications}. However, as far as we know, there are few ones devoted to the development of mechanisms that could lead to light right-handed neutrinos\cite{lightnuR}. Suppose a scenario where the left-handed neutrinos as well as the right-handed ones are all light particles. In a scenario like this, a challeging task to particle physics  would be to develop a seesaw mechanism in the framework of some extention of the standard model that could induce the small masses of these neutrinos. In this regard, an even more interesting scenario would be one where the explanation of the lightness of both left-handed and right-handed neutrino masses would have a common origin.

In this paper we consider a variant of the gauge models based in the $SU(3)_C \otimes SU(3)_L \otimes U(1)_N$(3-3-1) symmetry called 3-3-1 model with right-handed neutrinos(331$\nu$R)\cite{footpp} and adapt the type II seesaw mechanism in this framework. We proceed as in the implementation of the conventional type II seesaw mechanism in the standard model. Due to the structure of the 331$\nu$R, instead of a triplet, we have to add a sextet of scalars to its particle content. As our main result, we will show that this type II seesaw mechanism will induce small masses for the left-handed and right-handed neutrinos. Moreover, both neutrino masses have a common origin. As common origin we mean that both masses arise from the same Yukawa term, or better, the same set of Yukawa couplings are common for both neutrino masses. Another interesting point is that the mechanism can provide right-handed neutrinos much heavier than the left-handed ones. For example, for  typical values of the free parameters of the model we can obtain at least one right-handed neutrino with mass in the  KeV range. This is particularly interesting because right-handed neutrino with mass in this range is a viable candidate for the warm component of the dark matter existent in the universe.

This work is organizwed as follow. In Sec.~(\ref{sec2}) we present some aspects of the model relevant for the implementaion of the mechanism. In Sec.~(\ref{sec3}) we implement the mechansim in the model and present an ilustrative example. We finish this work with a summary in Sec.~(\ref{sec4}).

\section{Some aspects of the model}
\label{sec2}

In the 331$\nu$R the leptons come in triplet and singlets as follows\cite{footpp},
\begin{eqnarray}
f_{aL} = \left (
\begin{array}{c}
\nu_a \\
e_a \\
\nu^{c}_a
\end{array}
\right )_L\sim(1\,,\,3\,,\,-1/3)\,,\,\,\,e_{aR}\,\sim(1,1,-1),
 \end{eqnarray}
with $a=1,2,3$ representing the three known generations. We are
indicating the transformation under 3-3-1 after the similarity
sign, ``$\sim$''.  

In the gauge sector, the model recovers the standard gauge bosons  and disposes of five more other called  $V^{\pm}$, $U^0$, $U^{0 \dagger}$ and $Z^{\prime}$\cite{footpp}.

The scalar sector of the model is composed by three scalar triplets as follows\cite{footpp}
\begin{equation}
\chi=
\begin{pmatrix}
\chi^{0} \\
\chi^{-} \\
\chi'^{0}
\end{pmatrix}
,
\rho=
\begin{pmatrix}
\rho^{+} \\
\rho^{0} \\
\rho'^{+}
\end{pmatrix}
,
\eta=
\begin{pmatrix}
\eta^{0} \\
\eta^{-} \\
\eta'^{0}
\end{pmatrix}
,
\end{equation}
with $\eta$ and $\chi$ transforming as (1,3,$-\frac{1}{3}$) and $\rho$ transforming as (1,3,$-\frac{2}{3}$). Althoug the scalar content above involves  five neutral scalars, it is just necessary that $\eta^0\,,\,\rho^0$ and $\chi^{\prime 0}$ develop vaccum expectation value(VEV) for that all the particles of the model, with exception of the neutrinos, develop their correct mass terms. 

In the 331$\nu$R neutrino mass terms must, necessarily, involve the product  $\bar{f}_{L}f_{L}^{c}$. Considering its transformation  by the $SU(3)_L$ symmetry, $\bar{f}_{L}f_{L}^{c}=3^{*} \otimes 3^{*}=3 \oplus 6^{*}$, we see that, in order to generate mass terms for the neutrinos, we must couple  $\bar{f}_{L}f_{L}^{c}$ to an anti-triplet or to a sextet of scalars. Previous works showed that the first case leads to degenerate Dirac mass terms for the neutrinos\cite{footpp}. This case is not of interest for us here. In regard to the scalar sextet, recent works have employied it  to implement the type I seesaw mechanism into the model\cite{usesextet}. 

In this work we employ the scalar sextet to implement the type II seesaw mechanism into model\cite{typeIIintheminimalmodel}. The sextet we refer has the following scalar content,
\begin{equation}
S=
\dfrac{1}{\sqrt{2}}
\begin{pmatrix}
\Delta^{0} & \Delta^{-} & \Phi^{0} \\
\Delta^{-} & \Delta^{--} & \Phi^{-} \\
\Phi^{0} & \Phi^{-} & \sigma^{0}
\end{pmatrix}
\sim (1,6,\frac{-2}{3}).
\end{equation}
With the lepton triplet $f_L$  and the sextet $S$ we form the Yukawa coupling $\bar f_L f^C_L S$. The sextet $S$ presents three neutral scalars that may develop VEV. The VEV's of the neutral scalars  $\Delta^0$ and  $\sigma^0$ will lead to  Majorana mass terms for both the left-handed and right-handed neutrinos respectively, while the VEV of the neutral scalar  $\Phi^0$  will generate a Dirac mass term which mix the left-handed neutrinos with the  right-handed ones. In order to implement the type II seesaw mechanism the  Dirac mass terms must be avoided. For this we assume that  $\Phi^0$ does not develop VEV and impose the set of discrete symmetries $(\chi,\rho, e_{aR}) \rightarrow -(\chi,\rho, e_{aR})$. The discrete symmetry also helps in avoiding flavor changing neutral currents involving quarks and scalars and is important to obtain a simple potential. In regard to the Dirac mass terms, we think important to emphasize that, in avoiding them,  the left-handed neutrinos get decoupled of the right-handed ones. Thus, in this case, we could say that our right-handed neutrinos are, in fact, completly steriles.

The mechanism arises in the potential of the model and is communicated to the neutrinos through the Yukawa interaction $\bar f_L f^C_L S$. The essence of the mechanism is that lepton number is violated explictly through some terms in the potential. For this it is necessary to know the lepton number distribution of the scalars: $L(\eta^{\prime 0}\,,\,\sigma^0 \,,\,\rho^{\prime +})=-2$  and  $L(\chi^0\,,\, \chi^-\,,\,\Delta^0\,,\,\Delta^-\Delta^{--})=2$.  When $\Delta^0$ and $\sigma^0$ develop VEV's, automaticaly the left-handed neutrinos($\nu_L$)  and the right-handed ones($\nu_R$) both  develop  Majorana mass terms.  The masses of $\nu_L$ and  of $\nu_R$ get proportional to $v_\Delta$  and  $v_\sigma$, respectively. The role of the type II seesaw mechanism  is to furnish tiny values for $v_\Delta$  and $v_\sigma$.

The most complete part of the potential that obeys the discrete symmetry discussed above and conserves lepton number is composed by the following terms, 
\begin{eqnarray}
V & = & \mu_{\chi}^{2}\chi^{2}+\mu_{\eta}^{2}\eta^{2}+\mu_{\rho}^{2}\rho^{2}+\lambda_{1}\chi^{4}+\lambda_{2}\eta^{4}+\lambda_{3}\rho^{4}+\lambda_{4}(\chi^{\dagger}\chi)(\eta^{\dagger}\eta) \nonumber\\
  & + & \lambda_{5}(\chi^{\dagger}\chi)(\rho^{\dagger}\rho)+\lambda_{6}(\eta^{\dagger}\eta)(\rho^{\dagger}\rho)+\lambda_{7}(\chi^{\dagger}\eta)(\eta^{\dagger}\chi)+\lambda_{8}(\chi^{\dagger}\rho)(\rho^{\dagger}\chi) \nonumber\\
  & + & \lambda_{9}(\eta^{\dagger}\rho)(\rho^{\dagger}\eta)+(\frac{f}{\sqrt{2}}\epsilon^{ijk}\eta_{i}\rho_{j}\chi_{k}+H.C)+\mu_{S}^{2}Tr(S^{\dagger}S) \nonumber\\
  & + & \lambda_{10}Tr(S^{\dagger}S)^{2}+\lambda_{11}[Tr(S^{\dagger}S)]^{2}+(\lambda_{12}\eta^{\dagger}\eta+\lambda_{13}\rho^{\dagger}\rho+\lambda_{14}\chi^{\dagger}\chi)Tr(S^{\dagger}S) \nonumber\\
  & + & \lambda_{15}(\epsilon^{ijk}\epsilon^{lmn}\rho_{n}\rho_{k}S_{li}S_{mj}+H.C)+\lambda_{16}(\chi^{\dagger}S)(S^{\dagger}\chi)+\lambda_{17}(\eta^{\dagger}S)(S^{\dagger}\eta) \nonumber\\
  & + & \lambda_{18}(\rho^{\dagger}S)(S^{\dagger}\rho),
\end{eqnarray}
while the other part  that violates explicitly the lepton number is composed by these terms,  
\begin{eqnarray}
V^{\prime} & = &
\lambda_{19}(\eta^{\dagger}\chi)(\eta^{\dagger}\chi)+(\frac{\lambda_{20}}{\sqrt{2}}\epsilon^{ijk}\eta_{m}^{*}S_{mi}\chi_{j}\rho_{k}+H.C)+(\frac{\lambda_{21}}{\sqrt{2}}\epsilon^{ijk}\chi_{m}^{*}S_{mi}\eta_{j}\rho_{k}+H.C)\nonumber\\
        & - & M_{1}\eta^{T}S^{\dagger}\eta-M_{2}\chi^{T}S^{\dagger}\chi.
\end{eqnarray}
We think we have presented all the aspects of the model that are relevant to the implementation of the type II seesaw mechanism, which we do next. 
\section{The implementation of the type II seesaw mechanism}
\label{sec3}
From the Yukawa interaction,
\begin{eqnarray}
	{\cal L}^Y_\nu = G_{ab}\bar f_{aL} S f^C _{bL} + \mbox{H.c},
	\label{yukawaforneutrinos}
\end{eqnarray}
when $\Delta^0$  and $\sigma^0$ both develop VEV,  the left-handed and the right-handed neutrinos develop the folowing mass terms, 
\begin{eqnarray}
	{\cal L}^Y_\nu = G_{ab}v_\Delta\bar \nu_{aL}^C \nu_{bL}  + G_{ab}v_\sigma\bar \nu_{aR}^C \nu_{bR} .
	\label{preliminarymassmatrix}
\end{eqnarray}
We emphasize here that the mass terms of both neutrinos have as common origin the Yukawa interaction in Eq. (\ref{yukawaforneutrinos}). In practical terms this means that the same Yukawa couplings  $G_{ab}$ are common for the left-handed and right-handed neutrino masses. That's a very interesting result because when the masses of the left-handed neutrinos get measured directly, automatically the masses of the right-handed neutrinos will be predicted.

The role of the type II seesaw mechanism here is to provide tiny values for  $v_\Delta$  and $v_\sigma$.  This is achieved from the minimum condition of the  potential $V+V^{\prime}$. For this it is necessary to select which of the eight neutral scalars of the model  develop VEV. We already discussed why  $\Phi^0$ are not allowed here to develop VEV. In  the original version of the model only $\eta^0$, $\rho^0$  and $\chi^{\prime o}$ were allowed to develop VEV. The reason for this is to avoid flavor changing neutral currents involving  quarks and scalars. Thus we have that the simplest case is when only $\chi^{\prime 0}\,,\, \rho^{0}\,,\, \eta^{0}\,,\, \Delta^{0}\,,\,\sigma^{0} $  develop VEV. We assume this and shift these fields in the following way,
\begin{equation}
\chi'^{0}, \rho^{0}, \eta^{0}, \Delta^{0}, \sigma^{0} \rightarrow \dfrac{1}{\sqrt{2}}(v_{\chi', \rho, \eta, \Delta, \sigma}+R_{\chi', \rho, \eta, \Delta, \sigma}+iI_{\chi', \rho, \eta, \Delta, \sigma})
\label{shift}.
\end{equation}

Considering the shift of the neutral scalars in Eq. (\ref{shift}), the composite potential, $V+V^{\prime}$, provides the following set of minimum conditions,

\begin{eqnarray}
\mu_{\rho}^{2}&&+\lambda_{3}v_{\rho}^{2}+\frac{\lambda_{5}}{2}v_{\chi'}^{2}+\frac{\lambda_{6}}{2}v_{\eta}^{2}+\frac{fv_{\eta}v_{\chi'}}{2v_{\rho}}+\frac{\lambda_{13}}{4}(v_{\sigma}^{2}+v_{\Delta}^{2})+\nonumber\\
  &   & \lambda_{15}v_{\Delta}v_{\sigma}-\frac{\lambda_{20}}{4}(\frac{v_{\eta}v_{\Delta}v_{\chi'}}{v_{\rho}})+\frac{\lambda_{21}}{4}(\frac{v_{\chi'}v_{\eta}v_{\sigma}}{v_{\rho}})=0,
  \nonumber \\
  \mu_{\eta}^{2}&&+\lambda_{2}v_{\eta}^{2}+\frac{\lambda_{4}}{2}v_{\chi'}^{2}+\frac{\lambda_{6}}{2}v_{\rho}^{2}+\frac{fv_{\chi'}v_{\rho}}{2v_{\eta}}+\frac{\lambda_{12}}{4}(v_{\sigma}^{2}+v_{\Delta}^{2})+\nonumber\\
  &   & \frac{\lambda_{17}}{4}v_{\Delta}^{2}-M_{1}v_{\Delta}-\frac{\lambda_{20}}{4}(\frac{v_{\Delta}v_{\rho}v_{\chi'}}{v_{\eta}})+\frac{\lambda_{21}}{4}(\frac{v_{\chi'}v_{\rho}v_{\sigma}}{v_{\eta}})=0,\nonumber \\
  \mu_{\chi}^{2}&&+\lambda_{1}v_{\chi'}^{2}+\frac{\lambda_{4}}{2}v_{\eta}^{2}+\frac{\lambda_{5}}{2}v_{\rho}^{2}+\frac{fv_{\eta}v_{\rho}}{2v_{\chi'}}+\frac{\lambda_{14}}{4}(v_{\Delta}^{2}+v_{\sigma}^{2})+\nonumber\\
  &   & \frac{\lambda_{16}}{4}v_{\sigma}^{2}-M_{2}v_{\sigma}-\frac{\lambda_{20}}{4}(\frac{v_{\eta}v_{\rho}v_{\Delta}}{v_{\chi'}})+\frac{\lambda_{21}}{4}(\frac{v_{\eta}v_{\rho}v_{\sigma}}{v_{\chi'}})=0,\nonumber \\
  \mu_{S}^{2}&&+\frac{\lambda_{10}}{2}v_{\Delta}^{2}+\frac{\lambda_{11}}{2}(v_{\sigma}^{2}+v_{\Delta}^{2})+\dfrac{\lambda_{12}}{2}v_{\eta}^{2}+\dfrac{\lambda_{13}}{2}v_{\rho}^{2}+\dfrac{\lambda_{14}}{2}v_{\chi'}^{2}+\lambda_{15}\frac{v_{\rho}^{2}v_{\sigma}}{v_{\Delta}}+
  \nonumber\\
 && \dfrac{\lambda_{17}}{2}v_{\eta}^{2}-\frac{\lambda_{20}}{2}(\frac{v_{\eta}v_{\rho}v_{\chi'}}{v_{\Delta}})-M_{1}\frac{v_{\eta}^{2}}{v_{\Delta}}=0,
  \nonumber \\ \mu_{S}^{2}&&+\frac{\lambda_{10}}{2}v_{\sigma}^{2}+\frac{\lambda_{11}}{2}(v_{\sigma}^{2}+v_{\Delta}^{2})+\dfrac{\lambda_{12}}{2}v_{\eta}^{2}+\dfrac{\lambda_{13}}{2}v_{\rho}^{2}+\dfrac{\lambda_{14}}{2}v_{\chi'}^{2}+\lambda_{15}\frac{v_{\rho}^{2}v_{\Delta}}{v_{\sigma}}+\nonumber\\
  &   & \dfrac{\lambda_{16}}{2}v_{\chi'}^{2}-M_{2}\frac{v_{\chi'}^{2}}{v_{\sigma}}+\frac{\lambda_{21}}{2}(\frac{v_{\eta}v_{\rho}v_{\chi'}}{v_{\sigma}})=0.
  \label{minimum}
\end{eqnarray}

In any conventional seesaw mechanism,  the masses of the particles inherent of the mechanism and the energy scale associated to the  violation of the lepton number both must  lie in the GUT range. Bringing this to our mechanism, we have that the masses of the scalars that compose the sextet, $\mu_S$, and the energy scale $M_1$  and $M_2$ that appear in terms that violated explictly the lepton number both must lie in the GUT range which is around $10^{12}-10^{14}$GeV. For sake of simplicity, here we assume  $\mu_S\approx M_1 \approx M_2=M$.

As   $v_{\sigma\,,\,\Delta} << v_{\chi^{\prime}, \rho, \eta}$  and $v_{\chi^{\prime}, \rho, \eta} << M$, we see that the fourth expression in  Eq. (\ref{minimum}) provides $v_\Delta = \frac{v^2_\eta}{M}$  while the last one provides $v_\sigma = \frac{v^2_{\chi^{\prime}}}{M}$.   On substituting these expressions for the VEV's $v_\Delta $  and  $v_\sigma$ in Eq. (\ref{preliminarymassmatrix}), we obtain,
\begin{eqnarray}
	m_{\nu_L}=G\frac{v^2_{\eta}}{M}\,\,\,\,\,\,\,\,\,\,m_{\nu_R}= G\frac{v^2_{\chi^{\prime}}}{M}.
	\label{finalmassmatrix}
\end{eqnarray}
Notice that, the higher the $M$, the smaller the masses of $\nu_L$  and $\nu_R$. This is our type II seesaw mechanism where the masses of the neutrinos are surppressed by the high energy scale $M$. To go further, and make some predictions, there are no other way unless take fine-tunning of the free parameters involved in the mechanism. However, note that, apart from the Yukawa coupling $G_{ab}$, we have that $\frac{m_{\nu_R}}{m_{\nu_L}}=\frac{v^2_{\chi^{\prime}}}{v^2_{\eta}}$. For typical values of both VEV's, for example, $v_{\eta}=10^2$GeV  and $v_{\chi^{\prime}}=10^4$GeV we obtain $\frac{m_{\nu_R}}{m_{\nu_L}}=10^4$. Thus, for left-handed neutrinos of mass of order of $10^{-1}$eV  we may have right-handed neutrinos of mass of few KeV. This is an encouraging result because right-handed neutrinos with mass in this range  is a viable warm dark matter candidate\cite{dodelson}.

There are not enough experimental data in neutrino physics capable of fixing all the Yukawa couplings $G_{ab}$ that appear in the neutrino mass expressions above. Even the  solar and atmospheric neutrino oscillation experimental data are not able to do this. There is no better way of proceeding here than  choose a determined texture for the mass matrix $m_{\nu_L}$ whose diagonalization yield the correct neutrino mass squared difference and the maximal mixing angles involved in solar and atmospheric neutrino oscillation. This is obtained by assuming a determined set of values for the Yukawa couplings  $G_{ab}$. Having in mind that we look for a scenario where the right-handed neutrinos may come to be a viable candidate for dark matter, we than select a set of values for   $G_{ab}$  that leave the right-handed neutrinos as heavy as possible.

As an illustrative example, we take  $v_\eta=40$GeV, $M=10^{12}$GeV and the following set of values for Yukawa couplings: 
\begin{eqnarray}
	&&G_{11}=0.001924421313\,\,\,,\,\,\, G_{12}= 0.001837797437\,\,\,,\,\,\,  G_{13}= -0.001837797437\nonumber \\
&& G_{22}=0.01742997684\,\,\,,\,\,\,G_{23}=0.01382002316\,\,\,,\,\,\,G_{33}=0.01742997684.  
\label{yukawacoupling}
\end{eqnarray}
With this we  obtain the following texture for the mass matrix $m_{\nu_L}$,
\begin{eqnarray}
m_{\nu_L}=
\begin{pmatrix}
0.003079074101 & 0.002940475899 & -0.002940475899 \\
 0.002940475899& 0.02788796295 & 0.02211203705 \\
-0.002940475899& 0.02211203705 & 0.02788796295
\end{pmatrix}eV.
\label{finalmnuL}
\end{eqnarray}
The diagonalization of this mass matrix yields the following left-handed neutrino masses
\begin{equation}
m_{1}\approx 5.5\times 10^{-5}eV, \,\,\, m_{2}\approx 8.8\times10^{-3}eV, \,\,\, m_{3}\approx 5.0\times10^{-2}eV.
\end{equation}
This predictions for the left-handed neutrino masses imply the following values for the neutrino mass squared differences,
\begin{equation}
\Delta m_{21}^{2}=7.7 \times 10^{-5}eV^{2}, \,\,\,
\Delta m_{32}^2=2.4 \times 10^{-3}eV^{2}.
\end{equation}

Moreover, the  mass matrix above is diagonalized by the following mixing matrix,
\begin{equation}
U=
\begin{pmatrix}
0.809 & 0.588 & 0 \\
-0.416 & 0.572 & 0.707 \\
0.416 & -0.572 & 0.707
\end{pmatrix},\label{mixingmatrixnuL}
\end{equation}
The matrix in Eq.~(\ref{mixingmatrixnuL}) can be parametrized in terms of mixing angles {\it a la} Cabibbo-Kobayashi-Maskawa(CKM) parameterization~\cite{CKM} and is reproduced if we take: $ \theta_{12}=36^o$, $\theta_{23}=45^o$  and $ \theta_{13}=0$.  Thus, such mixing angles  together with the above neutrino mass squared differences, explain both the solar and atmospheric neutrino oscillations according to the current data\cite{kayser}.

With the Yukawa couplings given in eq. (\ref{yukawacoupling}), and taking $v_{\chi^{\prime}}=10^4$GeV,  we obtain the  following texture for the mass matrix $m_{\nu_R}$,
\begin{equation}
m_{\nu_R}=
\begin{pmatrix}
192.4421313 & 183.7797437 & -183.7797437 \\
183.7797437  & 1742.997684 & 1382.002316 \\
-183.7797437 & 1382.002316 & 1742.997684
\end{pmatrix}eV.
\label{mnuRtexture}
\end{equation}
On Diagonalizing this mass matrix we obtain the following  predictions for the right-handed neutrino masses,
\begin{equation}
m_{4}\approx 3.5eV, \,\,\, m_{5}\approx 550eV, \,\,\, m_{6}\approx 3.2KeV.
\label{predictionmassnuR}
\end{equation}

In this illustrative example our  heavier right-handed neutrino has mass of $3.2$KeV, and for sterile neutrino be a warm dark matter candidate its mass must lie in the range $0.3\mbox{KeV}< m_{\nu_R}<3.5$KeV where the lower bound is obtained by tremaine-Gunn bounds\cite{tremainebound} and the upper bound is obtained by radiative decays of sterile neutrinos in dark matter halos limited by X-ray observations\cite{xraybound}.

For we verify if this neutrino is  a viable dark matter candidate, first thing to do is to check if it is stable. For a neutrino with mass in the KeV range its unique possibility of decaying is in other lighter neutrinos. The interaction that engender such decay is given by,
\begin{equation}
\frac{g}{\sqrt{2}}\overline{(\nu_{R})^{c}}\gamma^{\mu}\nu_{L}U_{\mu}^{0\dag}+ H.C.
\label{nuRnulU}
\end{equation}
Note that the mass matrices for $m_{\nu_L}$  and $m_{\nu_R}$ given in Eq. (\ref{finalmassmatrix}) are both diagonalized by the same mixing matrix $U$. In consequence, the interaction above is always diagonal in any basis. In view of this, we can have the following neutrino decay channel $\nu_6 \rightarrow \nu_3 \, \nu_4 \, \nu_1$. For this channel we obtain the following decay width:
\begin{equation}
\Gamma=\frac{G_{F}^{2} m_6^{5} m_W^{4}}{192 \pi^{3} m_{U}^{4}}.
\label{decaywidth}
\end{equation}
The mass of the gauge boson $U^0$  is given by $m^2_U=\frac{g^2}{4}(v^2_\eta +v^2_{\chi^{\prime}})$. For the values of VEV's assumed above we obtain $m_U=3250$GeV, which implies  $\nu_6$ of $\tau=2.3\times 10^{23}$s for the lifetime of the neutrino $\nu_6$. Based on the WMAP best fit\cite{wmap}, the age of the universe is $\tau_0 =2.1\times 10^{17}$s. This means that $\nu_6$  is stable in face of the present age of the universe. Stable right-handed neutrino(for now on sterile neutrino) with mass in the KeV range is warm dark matter candidate\cite{dodelson,implications}. However for a  sterile neutrino be a viable dark matter candidate it has to satisfy all the cosmologial and astrophysical constraints. All those constraints depend on the mechanism of production of these neutrinos in the early universe\cite{astro-cosmo,dodelson,bode}. The most popular mechanism  is through resonant or non-resonant production via active-sterile neutrino mixing\cite{dodelson}. All the current constraint on sterile neutrino were derived using one of these mechanism of production.  As our sterile neutrino  does not mix with the active ones, we conclude that all the current cosmological and astrophysical  constraints  can not be applied to our sterile neutrino. All the other mechanisms of production of sterile neutrinos as decay\cite{decay} and scattering mecanisms\cite{khalil} are model dependent.  In our model, we think that the most efficient  mechanism of production capable of generating right amount of  sterile neutrinos in the early universe is the scattering production through reaction like $e^+ + e^- \rightarrow \bar \nu_6+\nu_6 $ intermediated by the charged gauge boson $V^{\pm}$ according to the interaction $\frac{g}{\sqrt{2}}\overline{(\nu_{R})^{c}}\gamma^{\mu}e^-_{L}V^+_\mu$. Thus, it seems that, in view of this mechanism of production of $\nu_6$, all the cosmological implications of this neutrino should be revisited. This will be explored in future works.
\section{Summary}
\label{sec4}

In this work we adapted the type II seesaw mechanism to the 3-3-1 model with right-handd neutrinos. We proceeded as in the implementation of the conventional mechanism to the standard model. Our major results are these:  the mechanism is  able of generating small masses simultaneously for the left-handed and right-handed neutrinos and that both masses have their origin in a common Yukawa coupling.  In addition, we obtained that at least one sterile right-handed neutrinos gain mass in the KeV range which turns it a viable candidate for the warm component of the dark matter existent in the universe. 
\acknowledgments
This work was supported by Conselho Nacional de Pesquisa e
Desenvolvimento Cient\'{i}fico- CNPq(HD,CASP) and Coordena\c c\~ao de Aperfei\c coamento de Pessoal de N\'{i}vel Superior - capes(DC).


\end{document}